\documentclass[12pt]{article}

\usepackage{amsmath}
\usepackage{wasysym}
\usepackage{graphicx}
\usepackage[stable]{footmisc}

\begin{document}

\thispagestyle{empty}
\def\thefootnote{\fnsymbol{footnote}}

\hfill   LBNL-59221

\hfill   UCB-PTH-05/40

\hfill	hep-th/0512199

\hfill   December 15, 2005

\begin{center}

\vspace{18pt}
{\bf Cosmological Consequences of String Axions}\footnote{This
work was supported in part by the
Director, Office of Science, Office of High Energy and Nuclear
Physics, Division of High Energy Physics of the U.S. Department of
Energy under Contract DE-AC02-05CH11231, in part by the National
Science Foundation under grant PHY-0098840.}
\vspace{18pt}

Ben Kain\footnote{E-Mail: {\texttt bkain@berkeley.edu}}
\vskip .01in
{\em Department of Physics, University of California 
and \\ Theoretical Physics Group, Bldg. 50A5104,
Lawrence Berkeley National Laboratory \\ Berkeley,
CA 94720 USA}

\vskip .03in
\vspace{18pt}

\end{center}

\begin{abstract} 

\noindent Axion fluctuations generated during inflation lead to isocurvature and non-Gaussian temperature fluctuations in the cosmic microwave background radiation.  Following a previous analysis for the model independent string axion we consider the consequences of a measurement of these fluctuations for two additional string axions.  We do so independent of any cosmological assumptions.  The first axion has been shown to solve the strong CP problem for most compactifications of the heterotic string while the second axion, which does not solve the strong CP problem, obeys a mass formula which is independent of the axion scale.  We find that if gravitational waves interpreted as arising from inflation are observed by the PLANCK polarimetry experiment with a Hubble constant during inflation of $H_\textnormal{inf} \apprge 10^{13}$ GeV the existence of the first axion is ruled out and the second axion cannot obey the scale independent mass formula.  In an appendix we quantitatively justify the often held assumption that temperature corrections to the zero temperature QCD axion mass may be ignored for temperatures $T\apprle\Lambda_\textnormal{QCD}$.

\end{abstract}

\def\thefootnote{\arabic{footnote}}\setcounter{footnote}{0}

\newpage
\thispagestyle{empty}

\mbox{ }

\vskip 1in

\begin{center}
{\bf Disclaimer}
\end{center}

\vskip .2in

\begin{scriptsize}
\begin{quotation}
\noindent This document was prepared as an account of work sponsored by the United States Government. Neither the United States Government nor any agency thereof, nor The Regents of the University of California, nor any of their employees, makes any warranty, express or implied, or assumes any legal liability or responsibility for the accuracy, completeness, or usefulness of any information, apparatus, product, or process disclosed, or represents that its use would not infringe privately owned rights. Reference herein to any specific commercial products process, or service by its trade name, trademark, manufacturer, or otherwise, does not necessarily constitute or imply its endorsement, recommendation, or favoring by the United States Government or any agency thereof, or The Regents of the University of California. The views and opinions of authors expressed herein do not necessarily state or reflect those of the United States Government or any agency thereof, or The Regents of the University of California and shall not be used for advertising or product endorsement purposes.
\end{quotation}
\end{scriptsize}

\vskip 2in

\begin{center}
\begin{small}
{\it Lawrence Berkeley Laboratory is an equal opportunity employer.}
\end{small}
\end{center}

\newpage
\setcounter{page}{1}

\section{Introduction}
\label{Intro}
It has been known for some time now that compactifications of string theory contain axion-like fields \cite{wi85}, while it has been an open question whether these string axions could be the QCD axion \cite{pqww}.  To date, axions have never been detected, and it seems a practical impossibility, due to their large axion scale,\footnote{One of the string axions that we will be considering, the BGW-2C axion (see section \ref{bgw2c}), does not solve the strong CP problem and therefore is not a Peccei-Quinn/QCD axion.  For this reason we will always refer to the axion coupling constant, $f_a$, as the ``axion scale" and not, say, the ``Peccei-Quinn scale."} that a string axion could be detected in the laboratory.  However, axions give rise to isocurvature \cite{st} and non-Gaussian \cite{ls} temperature fluctuations in the cosmic microwave background (CMB) radiation which may be measurable in the near future, even for a large axion scale.  If future experiments are able to measure the isocurvature and non-Gaussian components of CMB temperature fluctuations they could bound the axion scale and possibily rule out string axions as the QCD axion.  String axions are then one of the few known features of string theory that may be experimentally probed in the near future.

Recently P. Fox, A. Pierce and S. Thomas (FPT) \cite{fpt} analyzed the possibility that upcoming cosmological experiments sensitive to the polarization of the CMB (such as the PLANCK polarimetry experiment \cite{PLANCK}) may be able to probe a string axion by measuring the isocurvature and non-Gaussian components of the CMB.  For the string axion, FPT used the traditional model independent axion \cite{wi85}, and concluded that if gravitational waves interpreted as arising from inflation are observed by the PLANCK polarimetry experiment with a Hubble constant during inflation of $H_\textnormal{inf} \apprge 10^{13}$ GeV, then the model independent string axion cannot be the QCD axion.  

Here, we shall repeat their analysis for three different string axions.  Following FPT we do so independent of any cosmological assumptions, allowing for the widest possible range of cosmological conditions within allowed constraints.  Thus, our results are valid for any allowed cosmological scenario.  For completeness, and the purpose of comparison, the first axion shall be the same axion used by FPT, the model independent (MI) axion \cite{wi85}.  The second two axions, the primary concern of this paper, follow from the Bin\'etruy-Gaillard-Wu (BGW) model \cite{bgw1, bgw2}, a class of effective supergravity theories derived from the weakly coupled heterotic string.  These two axions are the BGW-QCD axion \cite{gk}, which for most compactifications of the heterotic string solves the strong CP problem and is therefore a legitimate candidate for the QCD axion, and the BGW-two condensate (BGW-2C) axion \cite{bgw2, bg}, which does not solve the strong CP problem, but may still have cosmological significance.  These axions will be reviewed in the next section.

Our conclusion is no different than FPT.  We find that if gravitational waves interpreted as arising from inflation with a Hubble constant during inflation of $H_\textnormal{inf} \apprge 10^{13}$ GeV are observed by the PLANCK polarimetry experiment, the BGW-QCD axion cannot be the QCD axion.  This result depends critically on the fact that the axion scale for the BGW-QCD axion differs during inflation and condensation.  We also find for a positive measurement by PLANCK that the BGW-2C axion cannot obey a scale independent mass formula.

In the next section we review the properties of the axions we will be considering.  In section \ref{Relic} we present relic axion densities both with and without entropy dilution, the derivation of which is reviewed in Appendix \ref{apro}.  In section \ref{cmb} the isocurvature and non-Gaussian CMB temperature fluctuations are discussed, paving the way for the axion constraints in section \ref{Constraints}.  Section \ref{Conclusion} presents our conclusions.  In Appendix \ref{ltam} we quantitatively justify the often held assumption that the zero temperature QCD axion mass may be used for temperatures $T \apprle \Lambda_\textnormal{QCD}$ (an assumption that will be made when calculating relic axion densities in section \ref{Relic} and Appendix \ref{apro}).  As we are following the analysis of FPT, the reader is referred to their excellent article \cite{fpt}, and references therein, for background and details.

\section{Axions}
\label{Axions}
We consider three different string axions, the model independent (MI) axion \cite{wi85}, the BGW-QCD axion \cite{gk} and the BGW-two condensate (BGW-2C) axion \cite{bgw2, bg}.  Each will be described in this section, the important results being their mass and axion scale formulas, which are summarized in section \ref{sum}.  The MI axion was analyzed in \cite{fpt}, whose analysis we repeat in this paper for completeness and for comparison with the BGW axions, our primary concern.  The BGW-QCD and BGW-2C axions are derived from the BGW supergravity effective string model \cite{bgw1, bgw2}.

We will find that the axion scales are very large ($f_a\apprge10^{16}$ GeV).  Such large scales require the initial misalignment angle (see section \ref{Relic} and Appendix \ref{apro}) to be extremely small ($\theta_i\apprle10^{-3}$ \cite{fpt}), otherwise the relic axion density would be in contradiction with experiment (being larger than the observed dark matter density).  Barring some mechanism that can set the initial misalignment angle to zero (or to an extremely small value), we have a highly fine tuned requirement.  However, our purpose here does not require any such mechanism, as we are considering the axion without making any cosmological assumptions.  We will find, in fact, that this means taking the initial misalignment angle to be zero, as this leads to the weakest possible bounds (see section \ref{Relic}).  So although there is an obvious difficulty with such large axion scales, it will not affect our conclusions.

\subsection{Model Independent (MI) String Axion\footnote{In \cite{gk} the model independent string axion was referred to as the universal string axion.\label{fn}}}
\label{MI}
The massless spectrum of any superstring contains the graviton, two-form gauge potential, dilaton and their superpartners.  These fields are naturally formulated in terms of a linear superfield in a supergravity theory, which is dual to a formulation in terms of a chiral superfield \cite{bgg}.  The chiral superfield contains the scalar dilaton as well as a pseudoscalar, the MI axion.  For simplicity we present the chiral superfield formulation of the \textit{globally} supersymmetric Lagrangian (the supergravity generalization being straightforward \cite{bgg}).  The Langrangian in the chiral superfield formulation is given by \cite{wi85}
\begin{equation}
	{\cal L} =-M_p^2\int d^4\theta \ln(S+\bar{S}) +\frac{1}{4}\left(\int d^2\theta S W_a^{\alpha}W^a_\alpha + \textnormal{h.c.}\right),
	\label{chirallag}
\end{equation}
where $S$ is the chiral superfield, $W^\alpha_a$ is the superfield strength and $M_p=2.4\times10^{18}$ GeV is the reduced Planck mass.  Expanding the second term into component fields we find
\begin{equation}
	\frac{1}{4}\int d^2\theta S W^a_\alpha W_a^\alpha+ \textnormal{h.c} \supset -\frac{1}{8}(S+\bar{S})F^a_{\mu\nu} F_a^{\mu\nu} + \frac{i}{8}(S-\bar{S})F^a_{\mu\nu} \tilde{F}_a^{\mu\nu},
\end{equation}
where we have used the same symbol, $S$, for the lowest component of the chiral superfield, $F^a_{\mu\nu}$ is the Yang-Mills field strength and $\tilde{F}^a_{\mu\nu}=\frac{1}{2}\epsilon_{\mu\nu\rho\sigma}F^{a\rho\sigma}$ is the dual field strength.  We notice the dilaton as the real part of $S$ since it acts as the coupling constant for the Yang-Mills kinetic term and the axion as the imaginary part of $S$ since it has the standard axion coupling to the gauge field.  The QCD $\theta$ angle is then given in terms of this string axion by
\begin{equation}
	\theta=4\pi^2 i(S-\bar{S}),
\end{equation}
while the unified gauge coupling constant, $\alpha_U$, is given by 
\begin{equation}
	\alpha_U^{-1}=\frac{4\pi}{g_U^2}=2\pi(S+\bar{S}),
\end{equation}
and we take $g_U\sim O(1)$.

Expanding the kinetic term of (\ref{chirallag}) into component fields we find
\begin{equation}
	-M_p^2\int d^4\theta \ln(S+\bar{S}) \supset \frac{M_p^2}{(8\pi^2)^2}\frac{2}{(\alpha_U^{-1}/2\pi)^2} \frac{1}{2}\partial_\mu \theta \partial^\mu \theta = \frac{1}{2}\partial_\mu a \partial^\mu a,
\end{equation}
where $a$ is the axion.  The axion scale, $f_a$, is then given by
\begin{equation}
	f_a/N=\frac{a}{\theta}=\sqrt{2}\frac{\alpha_U}{4\pi}M_p\sim 10^{16} \textnormal{ GeV},
	\label{miscale}
\end{equation}
where $N$ is the axion anomaly coefficient.

For the MI axion we use (non-stringy) mass formulas obtained from QCD.  These are the same mass formulas used by FPT in their analysis \cite{fpt}.  As reviewed in Appendix \ref{apro} the axion mass is temperature dependent.  The zero temperature axion mass is calculated by considering chiral symmetry breaking in QCD (see Appendix \ref{ltam}).  One finds \cite{pqww, sr85},
\begin{equation}
	m_a=\frac{2\sqrt{z}}{1+z}\frac{f_\pi}{f_a/N}m_\pi=1.2\times10^{-9}\textnormal{ eV }\left(\frac{10^{16} \textnormal{ GeV}}{f_a/N}\right),
	\label{mimass}
\end{equation}
where $z=m_u/m_d=0.56$, $m_\pi=135$ MeV and $f_\pi=93$ MeV is the pion decay constant.  In Appendix \ref{ltam} we justify the use of the zero temperature axion mass for (nonzero) temperatures $T\apprle\Lambda_\textnormal{QCD}$.  The finite temperature mass for $T\apprge\Lambda_{\textnormal{QCD}}$ may be calculated by considering instantons at finite temperature \cite{gpy}.  One finds \cite{pww}
\begin{equation}
		m_a(T) \simeq 2.2\times10^{-11} \textnormal{ eV} \left(\frac{10^{16}\textnormal{ GeV}}{f_a/N}\right) \left(\frac{\Lambda_\textnormal{QCD}}{200\textnormal{ MeV}}\right)^{1/2}\left(\frac{\Lambda_\textnormal{QCD}}{T}\right)^4,
	\label{htm}
\end{equation}
which is accurate for $T\apprge\Lambda_{\textnormal{QCD}}$, but not for too much larger.\footnote{In the next section we will assume the axions are massless during inflation.  For this to be the case, all contributions to the axion mass during inflation must be negligibly small.  In particular, this includes higher dimensional operators which can contribute to the axion mass.  To prevent such possibilities, FPT assumes that the Peccei-Quinn symmetry remains unbroken in the early universe, which they consider to be a mild assumption \cite{fpt}.  For the BGW axions, the possibility of higher dimensional operators was considered in \cite{bg, gk}, where it was found, for most compactifications of the heterotic string, to be negligible.  Thus, we do not require such an assumption for the BGW axions, and will therefore make no further mention of it.\label{massfn}}

\subsection{BGW-QCD String Axion}
\label{bgwqcd}
It has been an open question whether the MI axion of the previous subsection could be the QCD axion.  This question was recently revisited \cite{gk} in the context of the BGW model \cite{bgw1, bgw2} where it was shown that for most compactifications of the heterotic string, the MI axion\footnote{See footnote \ref{fn}.} solves the strong CP problem and is therefore a legitimate candidate for the QCD axion.  It was also found to have a modified axion scale, and being now model dependent, we refer to this string axion by a different name, the BGW-QCD axion.  Our concern here is to see if cosmological observations could rule out the possibility of the BGW-QCD axion being the QCD axion.

The BGW model is an effective supergravity model derived from the weakly coupled heterotic string, in which local supersymmetry is broken by gaugino condensation in a hidden sector, and is formulated in terms of the linear superfield instead of the chiral superfield.  Whereas in the chiral superfield formulation determining which field is the axion is rather simple because of its coupling to the gauge fields, this is not the case in the linear superfield formulation, where determining the axion field can be subtle \cite{bgw2, bg}.

It is the two-form gauge potential in the linear superfield formulation that is related to the pseudoscalar field (the string axion) of the chiral superfield formulation.  In the BGW model, below the condensation scale, the (unconfined) superfield strengths for the hidden sector gauge groups must be replaced by gaugino condensate chiral superfields, leading also to a replacement of the linear superfield by a vector superfield and a constraint condition \cite{bgw1}.  The phase of the lowest component of the gaugino condensate superfields is then related to the string axion.

For a single hidden sector condensing (simple) gauge group, or for more than one hidden sector condensing gauge group but with identical $\beta$-function coefficients, the axion is massless \cite{bgw2}.  For the case of a single hidden sector condensing gauge group, however, there is an enlarged symmetry in the limit of one or more massless quarks.  Taking into account the breaking of this symmetry was shown, in fact, to induce a mass for the axion as well as increase the axion scale compared to that of the MI axion \cite{gk}.  For two light quarks, below the scale of supersymmetry breaking, the axion scale is given by
\begin{equation}
	f_a/N \simeq \sqrt{\frac{8}{3}}\frac{1}{|8\pi^2 b_c - 3|}M_p,
	\label{bgwscale}
\end{equation}
where $b_c$ is the $\beta$-function coefficient for the single hidden sector condensing gauge group.  There exists a point of enhanced symmetry at $b_c=3/8\pi^2$, leading to an increase in the axion scale and, as shown below, a decrease in the axion mass.  In a study of electroweak symmetry breaking \cite{gn00} the preferred range for $b_c$ was found to be $.03\leq b_c \leq.04$, which requires 
\begin{equation}
	f_a/N\apprge 6\times10^{18}\textnormal{ GeV},
	\label{bgwqcdscale}
\end{equation}
which are the values of $f_a/N$ we will consider for the BGW-QCD axion.  The preferred value for LSP dark matter was found to be $b_c=.036$ \cite{bn}.  The zero temperature mass formula was determined to be the same as (\ref{mimass}),
\begin{equation}
	m_a=\frac{2\sqrt{z}}{1+z}\frac{f_\pi}{f_a/N}m_\pi =1.2\times10^{-11}\textnormal{ eV }\left(\frac{10^{18} \textnormal{ GeV}}{f_a/N}\right).
	\label{qcdmass}
\end{equation}
As with the MI axion we use this mass formula for temperatures $T\apprle\Lambda_\textnormal{QCD}$ and use the high temperature QCD mass formula (\ref{htm}) for $T\apprge\Lambda_\textnormal{QCD}$.

Above the scale of supersymmetry breaking the axion scale and mass are modified.  If we take the supersymmetry breaking scale to be $\sim$ TeV or above, then, as shown in Appendix \ref{apro}, we will not be in need of the modified axion mass.  However, in the next section, we will be in need of the axion scale during inflation, which is expected to be well above the scale of supersymmetry breaking.  In this case, we must use
\begin{equation}
	f'_a/N\simeq \sqrt{\frac{2}{3}}\frac{1}{4\pi^2 b_c}M_p\simeq10^{18}\textnormal{ GeV}.
	\label{susyscale}
\end{equation}

\subsection{BGW-Two Condensate (BGW-2C) String Axion}
\label{bgw2c}
As mentioned above, for a single hidden sector condensing gauge group, or for more than one hidden sector condensing gauge group but with identical $\beta$-function coefficients, the axion is massless.  In this case there is a nonanomalous R-symmetry.  This symmetry is broken if there are condensing gauge groups with different $\beta$-function coefficients, leading to a nonzero axion mass.  Such an axion does not solve the strong CP problem and therefore is not a candidate for the QCD axion.  It may, however, have cosmological significance, which warrants its inclusion in this paper.

In the case of two condensates with coefficients $b_1 > b_2$, the MI axion scale is reduced by a factor of $\sqrt{6}/b_1$ \cite{2cscale}.  As mentioned in the previous subsection, the preferred range of values for $b_1=b_c$ is $.03\apprle b_1 \apprle .04$.  The MI axion scale was given in (\ref{miscale}) as $\sim10^{16}$ GeV, so
\begin{equation}
	f_a/N\sim\frac{\sqrt{6}}{b_1}\times10^{16}\textnormal{ GeV}\sim 6\,\textnormal{--}\,8\times10^{17} \textnormal{GeV}.
	\label{2cscale}
\end{equation}
For condensation scales $\Lambda_1 \gg \Lambda_2$, the BGW-2C axion mass is approximately given by \cite{bgw2}
\begin{equation}
	m_a \approx \frac{6\sqrt{2}\langle\ell\rangle}{b_2} (b_1-b_2)\left(\frac{\Lambda_2}{\Lambda_1}\right)^{3/2} m_{\frac{3}{2}},
\end{equation}
where $\ell$ is the dilaton field in the linear multiplet formulation and $m_{\frac{3}{2}}$ is the gravitino mass which we take to be $m_{\frac{3}{2}}\sim 1$ TeV.  The important thing to notice about this formula is that it is independent of the axion scale.  In the context of the weakly coupled heterotic string a viable scenario for local supersymmetry breaking occurs if the hidden sector condensation scale is given by $\Lambda_1\sim10^{13}$ GeV \cite{gn00}.  If we take $\langle\ell\rangle\sim 1$ \cite{bgw2}, $b_1,b_2\sim10^{-2}$, $b_1-b_2\sim10^{-3}$ and $\Lambda_2\apprge\Lambda_\textnormal{QCD} \sim 10^2$ MeV we find
\begin{equation}
	m_a\apprge8\times 10^{-10}\textnormal{ eV}.
	\label{2cmass}
\end{equation}
This contribution to the BGW-2C axion mass exists for all energies below $\Lambda_2$, while above $\Lambda_2$ it disappears.

The BGW-2C axion also receives a contribution to its mass from the QCD (zero temperature) axion mass formula used for both the MI and BGW-QCD axions, (\ref{mimass}) and (\ref{qcdmass}).  For the axion scale in (\ref{2cscale}) this gives a mass $m_a\sim 6\,\textnormal{--}\,8\times10^{-10}$ eV, which is on the same order as (\ref{2cmass}).  Also, as shown in Appendix \ref{apro}, the temperature at which this axion condenses is always on the same order as $\Lambda_2$.  If the condensation temperature were greater than $\Lambda_2$ then the BGW-2C axion would not obey a scale independent mass formula during production in the early universe.  In this case, the cosmological consequences of this axion would not differ much from the previous two axions.  We therefore assume that the mass formula (\ref{2cmass}) is valid for calculating relic densities at all temperatures and only consider the BGW-2C axion when (\ref{2cmass}) is the dominant contribution, and thus for masses
\begin{equation}
	m_a\apprge10^{-9}\textnormal{ eV}.
	\label{2cmass2}
\end{equation}

We mention again that this axion does not solve the strong CP problem and is not a candidate for the QCD axion, but may still be cosmologically interesting, in particular because its mass formula (\ref{2cmass2}) is scale independent.

\subsection{Summary of Formulas}
\label{sum}
In the previous subsections we found the following masses and axion scales:
\begin{alignat}{2}
	\textnormal{MI:} \quad & m_a=1.2\times10^{-9}\textnormal{ eV }\left(\frac{10^{16} \textnormal{ GeV}}{f_a/N}\right), &&f_a/N\sim10^{16}\textnormal{ GeV},\\
	\textnormal{BGW-QCD:} \quad & m_a=1.2\times10^{-11}\textnormal{ eV }\left(\frac{10^{18} \textnormal{ GeV}}{f_a/N}\right), &\quad&f_a/N\apprge10^{18}\textnormal{ GeV},\\
	\textnormal{BGW-2C:} \quad &m_a\apprge10^{-9} \textnormal{ eV}, &&f_a/N\sim 10^{18}\textnormal{ GeV},
\end{alignat}
where the masses are to be used for temperatures $T\apprle\Lambda_\textnormal{QCD}$, except in the case of the BGW-2C axion where its mass is to be used for all temperatures.  We see that the first two are inversely proportional to the axion scale while the third one is independent of it.  When deriving relic axion densities we shall therefore use the following mass formulas:
\begin{align}
	m_a&=\xi \left(\frac{1\textnormal{ GeV}}{f_a/N}\right),\label{sdam}\\
	m_a&=\zeta,\label{siam}
\end{align}
which we will refer to as the scale dependent and scale independent mass formulas, and where
\begin{equation}
	\xi_\textnormal{MI}=1.2\times10^{7}\textnormal{ eV}, \qquad \xi_\textnormal{BGW-QCD}=1.2\times 10^{7} \textnormal{ eV}, \qquad \zeta_\textnormal{BGW-2C}\apprge10^{-9}\textnormal{ eV}. \label{mnum}
\end{equation}

For temperatures $T\apprge\Lambda_\textnormal{QCD}$ we use the following mass formula for the MI and BGW-QCD axions:
\begin{equation}
	m_a(T) \simeq 2.2\times10^5 \textnormal{ eV} \left(\frac{1\textnormal{ GeV}}{f_a/N}\right) \left(\frac{\Lambda_\textnormal{QCD}}{200\textnormal{ MeV}}\right)^{1/2}\left(\frac{\Lambda_\textnormal{QCD}}{T}\right)^4.
	\label{htam}
\end{equation}

Above the scale of supersymmetry breaking the axion scale for the BGW-QCD axion is modified.  We take $f'_a$ to be the axion scale above the scale of supersymmetry breaking and define
\begin{equation}
	f_s\equiv\frac{f_a}{f'_a},
	\label{ms}
\end{equation}
where
\begin{equation}
	f_s^\textnormal{MI}=1,\qquad f_s^\textnormal{BGW-QCD}\simeq\frac{f_a/N}{10^{18}\textnormal{ GeV}},\qquad f_s^\textnormal{BGW-2C}=1.
	\label{msval}
\end{equation}

\section{Relic Axion Densities}
\label{Relic}
The natural starting point when discussing axion cosmology is calculating the relic axion density.  The details of this calculation \cite{pww} are by now well known (see, for example, \cite{kt, fpt}), so we merely present and discuss the results here, relegating the details to Appendix \ref{apro} (where we calculate relic axion densities for axions obeying both scale dependent (\ref{sdam}) and scale independent (\ref{siam}) mass formulas).  Special attention is paid to the \emph{minimal} relic axion density, which will be used in section \ref{Constraints} to construct axion constraints.  

There are a number of different possible outcomes for the relic axion density (see (\ref{aprelicla})--(\ref{dilfb})).  In the absence of entropy dilution, for the scale dependent axion mass (\ref{sdam}) and for the axion condensing, respectively, during $T\apprle\Lambda_\textnormal{QCD}$ or $T\apprge\Lambda_\textnormal{QCD}$, the relic axion density is given by
\begin{align}
	\Omega_a h^2 &\simeq 1.8\times 10^{5} \left(\frac{\xi}{10^7\textnormal{ eV}}\right)^{1/2} \left(\frac{f_a/N}{10^{17} \textnormal{ GeV}}\right)^{3/2} (\left<\theta_i\right>^2 + \sigma_\theta^2) f(\theta_i)^2\notag\\
	&\quad \textnormal{for } f_a/N \apprge 2.7\times10^{17}\textnormal{ GeV} \left(\frac{\xi}{10^7\textnormal{ eV}}\right),\textnormal{ or} \label{reliclta}\\
\Omega_a h^2 &\simeq 2.4\times 10^{5} \left(\frac{\xi}{10^7\textnormal{ eV}}\right)\left(\frac{f_a/N}{10^{17} \textnormal{ GeV}}\right)^{7/6} (\left<\theta_i\right>^2 + \sigma_\theta^2) f(\theta_i)^2\notag\\
	&\quad \textnormal{for } f_a/N \apprle 2\times10^{15}\textnormal{ GeV},\label{relicht}
\end{align}
while for the scale independent mass (\ref{siam}), still in the absence of entropy dilution and for any condensation temperature,
\begin{equation}
	\Omega_a h^2 \simeq 3.5\times 10^{7} \left(\frac{\zeta}{10^{-9}\textnormal{ eV}}\right)^{1/2} \left(\frac{f_a/N}{10^{18} \textnormal{ GeV}}\right)^2 (\left<\theta_i\right>^2 + \sigma_\theta^2) f(\theta_i)^2. \label{relicltb}
\end{equation}
In the presence of entropy dilution, for the scale dependent (\ref{sdam}) or scale independent (\ref{siam}) axion mass, the relic axion density is given by, respectively,
\begin{align}
	\Omega_a h^2 \simeq 3.3&\times10^{3}\left(\frac{T_\textnormal{RH}}{6\textnormal{ MeV}}\right) \left(\frac{f_a/N}{10^{17}\textnormal{ GeV}}\right)^2 (\left<\theta_i\right>^2 + \sigma_\theta^2) f(\theta_i)^2\notag\\
	 \textnormal{for }&f_a/N \apprle 3.0\times10^{20}\textnormal{ GeV}\left(\frac{\xi}{10^7\textnormal{ eV}}\right)\left(\frac{6\textnormal{ MeV}}{T_\textnormal{RH}}\right)^2, \label{relicdila}\\
	\textnormal{or } &\zeta \apprge 3.3\times10^{-14}\textnormal{ eV}\left(\frac{T_\textnormal{RH}}{6\textnormal{ MeV}}\right)^2.
	\label{relicdilb}
\end{align}
In (\ref{reliclta})--(\ref{relicdilb}) $h=H_0/(100\textnormal{ km s}^{-1}\textnormal{ Mpc}^{-1})$, $N$ is a model dependent number on the order of one, $f(\theta_i^2)$ is a correction for anharmonic effects of the axion potential (which for small $\theta_i$ is roughly one), $T_\textnormal{RH}$ is the reheating temperature for the late decaying particle that dilutes the axion condensate (and must satisfy $T_\textnormal{RH}\apprge 6\textnormal{ MeV}$), $\left<\theta_i\right>^2 + \sigma_\theta^2=\left<\theta_i^2\right>$ is the mean square of the initial misalignment angle, $\theta_i$, averaged over the universe, and we have used $\Lambda_\textnormal{QCD}=200$ MeV.

The relic axion densities are all written in terms of the quantity
\begin{equation}
	\left<\theta_i^2\right>=\left<\theta_i\right>^2 + \sigma_\theta^2.
	\label{ma1}
\end{equation}
In the absence of inflation, each causally connected region of the universe would have a different (randomly selected) initial misalignment angle in the range $-\pi<\theta_i<\pi$, and the rms value would be $\langle\theta_i^2\rangle=\pi^2/3$.  In an inflationary universe, however, string axions exist before inflation and so the observable universe would have a single, constant, initial misalignment angle, since there existed a time in the early universe when the entire observable universe was in causal contact.  In this case, to use the rms value would be to average over each inflationary region, which tells us nothing about our observable universe.  Thus, it would not be correct to use the rms value for an inflationary universe.

Since we are assuming an inflationary universe we now interpret all quantities to refer to the observable universe only.  Thus, $\theta_i$, the initial misalignment angle, takes a constant value over the entire observable universe, so $\left<\theta_i\right>=\theta_i$.  $\sigma_\theta$ is the standard deviation of the initial misalignment angle, but since the observable universe has a single initial misalignment angle, it should be zero.  Below we shall reinterpret $\sigma_\theta$ as a characterization of quantum fluctuations, which are nonzero, and so shall leave it in equations.

Axions induce isocurvature \cite{st} and non-Gaussian \cite{ls} components in CMB temperature fluctuations.  These effects, as explained in section \ref{cmb}, may be parametrized in terms of the square and cube of the relic axion density, and the weakest bounds will be given by the minimal relic axion density, which itself is given by zero initial misalignment angle.  For zero initial misalignment angle, the relic axion densities are determined purely by quantum fluctuations during inflation (characterized by $\sigma_\theta$).

All massless fields undergo such quantum (de Sitter) fluctuations during inflation, and as long as the axion is (effectively) massless in the early universe,\footnote{See footnote \ref{massfn}.} it will too \cite{tw}.  The quantum fluctuations can be written in terms of the Hubble constant during inflation, $H_\textnormal{inf}$, as \cite{tw}
\begin{equation}
	\sigma_\theta=\frac{H_\textnormal{inf}}{2\pi(f'_a/N)}=f_s\frac{H_\textnormal{inf}}{2\pi(f_a/N)},
	\label{thetaH}
\end{equation}
where $f'_a$ is the axion scale (above the scale of supersymmetry breaking and therefore) during inflation and $f_s\equiv f_a/f'_a$.  The current bound on $H_\textnormal{inf}$ is $H_\textnormal{inf}\apprle 2.8\times 10^{14}$ GeV \cite{WMAPH}.

As mentioned above, we will be in need of the minimal relic axion densities.  From (\ref{reliclta})--(\ref{relicdilb}) we see that this corresponds to vanishing initial misalignment angle.  Replacing $\sigma_\theta$ in favor of $H_\textnormal{inf}$ we find for the minimal relic axion densities, in the absence of entropy dilution, for the scale dependent axion mass (\ref{sdam}) and for the axion condensing, respectively, during $T\apprle\Lambda_\textnormal{QCD}$ or $T\apprge\Lambda_\textnormal{QCD}$,
\begin{align}
	\Omega_a^\textnormal{min} h^2 &\simeq 4.6\times10^{-5} \,f_s^2\left(\frac{\xi}{10^7\textnormal{ eV}}\right)^{1/2} \left(\frac{10^{17} \textnormal{ GeV}}{f_a/N}\right)^{1/2} \left(\frac{H_\textnormal{inf}}{10^{13}\textnormal{ GeV}}\right)^2\notag\\
	&\quad \textnormal{for } f_a/N \apprge 2.7\times10^{17}\textnormal{ GeV} \left(\frac{\xi}{10^{7}\textnormal{ eV}}\right),\textnormal{ or} \label{relminlta}\\
	\Omega_a^\textnormal{min} h^2 &\simeq 6.2\times10^{-5}\,f_s^2\left(\frac{\xi}{10^7\textnormal{ eV}}\right) \left(\frac{10^{17}\textnormal{ GeV}}{f_a/N}\right)^{5/6} \left(\frac{H_\textnormal{inf}}{10^{13}\textnormal{ GeV}}\right)^2\notag\\
	&\quad \textnormal{for } f_a/N \apprle 2\times10^{15}\textnormal{ GeV},\label{relminhta}
\end{align}
while for the scale independent mass (\ref{siam}), still in the absence of entropy dilution and for any condensation temperature,
\begin{align}
	\Omega_a^\textnormal{min} h^2 &\simeq 8.8\times10^{-5} \,f_s^2\left(\frac{\zeta}{10^{-9}\textnormal{ eV}}\right)^{1/2} \left(\frac{H_\textnormal{inf}}{10^{13}\textnormal{ GeV}}\right)^2. \label{relminltb}
\end{align}
In the presence of entropy dilution, in which the minimal relic axion density is given for maximum entropy dilution, $T_\textnormal{RH}\simeq 6$ MeV, for the scale dependent (\ref{sdam}) or scale independent (\ref{siam}) axion mass, the relic axion density is given by, respectively,
\begin{align}
	\Omega_a^\textnormal{min} h^2 \simeq 8.4&\times10^{-7}\,f_s^2 \left(\frac{H_\textnormal{inf}}{10^{13}\textnormal{ GeV}}\right)^2 \notag\\
	\quad \textnormal{for } &f_a/N \apprle 3.0\times10^{20}\textnormal{ GeV}\left(\frac{\xi}{10^7\textnormal{ eV}}\right)\label{dilmina}\\
	\textnormal{or }&\zeta \apprge 3.3\times10^{-14}\textnormal{ eV}.	\label{dilminb}
\end{align}
These will be used in section \ref{Constraints} to construct axion constraints.

\section{CMB Temperature Fluctuations}
\label{cmb}
Quantum, spatially-dependent perturbations of fields during inflation can lead to observable temperature fluctuations in the CMB.  For example, inflationary perturbations of the inflaton, which is an example of an adiabatic perturbation, leads to adiabatic fluctuations in the CMB.  Adiabatic perturbations are characterized by fluctuations in the total energy density, $\delta\rho\neq0$, without fluctuations in the local equation of state, $\delta(n_i/s)=0$.  Our concern, however, will be with inflationary perturbations of the axion, which is an example of an isocurvature perturbation, and leads to isocurvature fluctuations in the CMB \cite{st}.  Isocurvature perturbations, characterized oppositely to adiabatic perturbations, correspond to fluctuations in the equation of state, $\delta(n_i/s)=0$, without fluctuations in the energy density, $\delta\rho\neq0$.  It is these isocurvature fluctuations, which contain the non-Gaussian fluctuations, that will lead to axion constraints.

We assume that the axion is the only contributor to isocurvature perturbations of the CMB (all other fields undergoing adiabatic perturbations) since this gives the weakest bounds.  Assuming otherwise would strengthen bounds by decreasing the axion isocurvature contribution for a (fixed) measured total isocurvature component of the CMB.  

We repeat here the derivation found in \cite{fpt} (to which the reader may look for details).  Isocurvature fluctuations of the axion may be characterized by
\begin{equation}
	S_a\equiv\frac{\delta(n_a/s)}{n_a/s}=\frac{\delta n_a}{n_a}-3\frac{\delta T}{T},\quad s\propto T^3,
	\label{Sa1}
\end{equation}
where we shall define exactly what we mean by $\delta$ below.  By considering the fluctuation of the total energy density, it may be shown that initially, just after the axion condensate is formed, the isocurvature temperature fluctuations, $\delta T/T$ in (\ref{Sa1}), satisfy $(\delta T/T)_\textnormal{init} \ll (\delta n_a/n_a)_\textnormal{init}$.  As $n_a/s$, and hence $S_a$, is conserved, it follows that
\begin{equation}
	S_a\simeq \left(\frac{\delta n_a}{n_a}\right)_\textnormal{init}.
	\label{Sa2}
\end{equation}
 
$S_a$ may now be written in terms of the axion field, or equivalently, the $\theta$ angle.  Since we are assuming an inflationary universe, we consider only our observable universe.  In this case, the initial misalignment angle, $\theta_i$, is a constant (over the observable universe) and we define $\delta\theta$ to be the spatially-dependent quantum fluctuation and $\theta\equiv\theta_i+\delta\theta$.  Brackets $\langle \rangle$ will denote averaging over the observable universe, so $\left<\theta_i\right>=\theta_i$ since it is constant, and we take the fluctuation, $\delta\theta$, to be Gaussian distributed with zero mean, so $\left<\delta\theta\right>=0$.

We are now in a position to define what we mean by $\delta$ in (\ref{Sa1}) and (\ref{Sa2}).  For concreteness, we do so specifically for $\delta n_a$.  Since we are dealing with spatially dependent quantum fluctuations of the axion, and thus of the axion number density, $n_a$ is taken to be the average value of the axion number density (and is the actual number density in the limit of vanishing quantum fluctuations), averaged over the observable universe.  $\delta n_a$ is then defined as the difference between the actual number density at some point in space and the average value,  $n_a$.  To express these defintions in terms of the $\theta$ angle, we note from (\ref{nd}) that $n_a \propto \left<\theta^2\right>$, and so
\begin{equation}
	 \delta n_a \propto \theta^2 - \left<\theta^2\right>.
\end{equation}
Along with $\left<\theta^2\right> = \left<\theta_i^2\right>+\left<(\delta\theta)^2\right> = \theta_i^2 + \sigma_\theta^2$,
which defines $\sigma_\theta^2=\left<\theta^2\right>-\left<\theta\right>^2=\left<(\delta\theta)^2\right>$ as a characterization of the average quantum fluctuation of $\theta$, (\ref{Sa2}) may be written in terms of the $\theta$ angle as
\begin{equation}
	S_a\simeq\left(\frac{\delta n_a}{n_a}\right)_\textnormal{init}=\frac{(\theta_i+\delta\theta)^2 - \langle\theta^2\rangle}{\langle\theta^2\rangle}=\frac{2\theta_i\delta\theta+(\delta\theta)^2 - \sigma_\theta^2}{\theta_i^2+\sigma_\theta^2},
\end{equation}
from which we shall need
\begin{equation}
	\langle S_a^2\rangle \simeq 2\sigma_\theta^2\frac{2\theta_i^2+\sigma_\theta^2}{(\theta_i^2 + \sigma_\theta^2)^2}, \qquad \langle S_a^3\rangle \simeq 8\sigma_\theta^4 \frac{3\theta_i^2+\sigma_\theta^2}{(\theta_i^2 + \sigma_\theta^2)^3},
\end{equation}
where we have used $\left<(\delta\theta)^4\right>=3\sigma_\theta^4$ and $\left<(\delta\theta)^6\right>=15\sigma_\theta^6$, following from $\delta\theta$ being Gaussian distributed.

Finally, the CMB temperature fluctuations induced by isocurvature perturbations from an axion are given by (see \cite{fpt} for details)
\begin{equation}
	\left(\frac{\delta T}{T}\right)_\textnormal{iso} \simeq -\frac{6}{15} \frac{\Omega_a}{\Omega_m} S_a  \simeq -\frac{6}{15} \frac{\Omega_a}{\Omega_m} \frac{2\theta_i\delta\theta+(\delta\theta)^2 - \sigma_\theta^2}{\theta_i^2+\sigma_\theta^2},
	\label{Tiso}
	\end{equation}
where $\Omega_m$ is the density of all non-relativistic matter and $\Omega_a$ is the relic axion density presented in the previous section.  This equation takes into account Sachs-Wolfe contributions \cite{sw} which are red shifting perturbations by matter to the CMB photons while they are on their way to the detector.  From WMAP $\Omega_m h^2 = .135^{+.008}_{-.009}$ \cite{WMAP} for a $\Lambda$CDM cosmology.\\

\centerline{\textbf{Isocurvature}}

The observable effects of isocurvature fluctuations are conveniently parametrized by
\begin{equation}
	\alpha \equiv \frac{\left<(\delta T/T)^2_\textnormal{iso}\right>}{\left<(\delta T/T)^2_\textnormal{tot}\right>} \simeq 2\left(\frac{6}{15}\right)^2 \frac{(\Omega_a/\Omega_m)^2}{\left<(\delta T/T)^2_\textnormal{tot}\right>} \sigma_\theta^2\frac{2\theta_i^2+\sigma_\theta^2}{(\theta_i^2 + \sigma_\theta^2)^2},
	\label{alpha}
\end{equation}
where COBE has measured $\left< (\delta T/T)^2_\textnormal{tot}\right>^{1/2}\simeq1.1\times 10^{-5}$ \cite{COBE} and a current conservative bound for $\alpha$ is $\alpha \apprle .4$ \cite{alpha}.

In the next section we will rewrite this equation in terms of $H_\textnormal{inf}$ and the minimal relic axion densities of the previous section, thereby obtaining a lower bound on $\alpha$.\\

\centerline{\textbf{Non-Gaussianity}}

Although the axion experiences Gaussian fluctuations ($\delta\theta$ was taken to be Gaussian), it induces both Gaussian and non-Gaussian fluctuations in the CMB \cite{ls}.  The first term on the right hand side of (\ref{Tiso}), proportional to $2\theta_i\delta\theta$, is obviously Gaussian distributed, but the remaining term, proportional to $(\delta\theta)^2 - \sigma_\theta^2$, is not.  This is the axion induced non-Gaussian fluctuation in the CMB.  
	
Gaussian fluctuations with zero mean are completely defined in terms of two point functions since all odd point functions vanish and all even point functions are products of two point functions.  Thus, a non-vanishing three point function may act as a measurement of non-Gaussianity.  The observable effects of non-Gaussian fluctuations induced by axions are then conveniently parametrized by the dimensionless skewness,
\begin{equation}
	S_{3,\textnormal{iso}} \equiv \frac{\left\langle(\delta T/T)_\textnormal{iso}^3\right\rangle }{\left\langle(\delta T/T)_\textnormal{tot}^2\right\rangle^{3/2}} \simeq -8\left(\frac{6}{15}\right)^3\frac{(\Omega_a/\Omega_m)^3}{\left\langle(\delta T/T)^2_\textnormal{tot}\right\rangle^{3/2}} \sigma_\theta^4\frac{3\theta_i^2+\sigma_\theta^2}{(\theta_i^2+\sigma_\theta^2)^3}.
	\label{s3}
\end{equation}
Note that we have assumed that the axion is the only source of non-Gaussianity,\footnote{In certain scenarios, the axion cannot be the dominant source of non-Gaussianity \cite{bl}.  This does not affect our conclusions since additional sources could only strengthen bounds.} as this leads to the weakest bounds.  Assuming otherwise would strengthen bounds by decreasing the axion contribution to non-Gaussianity for a (fixed) measured total non-Gaussian component of the CMB.  COBE has measured $-S_{3,\textnormal{iso}}\apprle.06$ \cite{COBE} .

In the next section we will rewrite this equation in terms of $H_\textnormal{inf}$ and the minimal relic axion densities of the previous section, thereby obtaining a lower bound on $-S_{3,\textnormal{iso}}$.

\section{Axion Constraints}
\label{Constraints}
We may now use our results to constrain the axion.  As mentioned in the introduction, we shall do so under the minimal possible assumptions, leading to the weakest possible bounds.  

The axion induces isocurvature and non-Gaussian components in CMB temperature fluctuations as described by (\ref{alpha}) and (\ref{s3}), respectively.  A look at these equations tells us that $\alpha$ and $-S_3$ are minimized for minimal relic axion density, $\Omega_a^\textnormal{min}$.  We found in section \ref{Relic} that minimal relic axion density corresponds to vanishing initial misalignment angle.  Thus we may use the fact that
\begin{equation}
\begin{aligned}
	\alpha &\apprge 2\left(\frac{6}{15}\right)^2 \frac{(\Omega_a^\textnormal{min}/\Omega_m)^2}{\left<(\delta T/T)^2_\textnormal{tot}\right>} \\
	-S_{3,\textnormal{iso}} &\apprge  8\left(\frac{6}{15}\right)^3\frac{(\Omega_a^\textnormal{min}/\Omega_m)^3}{\left\langle(\delta T/T)^2_\textnormal{tot}\right\rangle^{3/2}}
\end{aligned}
\end{equation}
to constrain the axion.  The minimal relic axion densities were given in (\ref{relminlta})--(\ref{dilminb}).  Plugging these in and rearranging, we find, in the absence of entropy dilution, for the scale dependent axion mass (\ref{sdam}) and for the axion condensing during $T\apprle\Lambda_\textnormal{QCD}$,
\begin{equation}
\left.\begin{aligned}
	f_a/N &\apprge \frac{3.1\times10^{19}\textnormal{ GeV}}{\alpha}f_s^4\left(\frac{\xi}{10^7\textnormal{ eV}}\right) \left(\frac{H_\textnormal{inf}}{10^{13}\textnormal{ GeV}}\right)^4 \\
	f_a/N &\apprge \frac{6.1\times10^{19}\textnormal{ GeV}}{(-S_{3,\textnormal{iso}})^{2/3}} f_s^4\left(\frac{\xi}{10^7\textnormal{ eV}}\right) \left(\frac{H_\textnormal{inf}}{10^{13}\textnormal{ GeV}}\right)^4
\end{aligned} \right\}
\left.\begin{aligned}
\textnormal{for } &f_a/N \apprge \\
&2.7\times10^{17}\textnormal{ GeV} \left(\frac{\xi}{10^7\textnormal{ eV}}\right),
\end{aligned}\right.
\label{ltconsa}
\end{equation}
while for the axion condensing during $T\apprge\Lambda_\textnormal{QCD}$,
\begin{equation}
\left.\begin{aligned}
	f_a/N &\apprge \frac{4.4\times10^{18}\textnormal{ GeV}}{\alpha^{3/5}}f_s^{12/5}\left(\frac{\xi}{10^7\textnormal{ eV}}\right)^{6/5} \left(\frac{H_\textnormal{inf}}{10^{13}\textnormal{ GeV}}\right)^{12/5} \\
	f_a/N &\apprge \frac{6.6\times10^{18}\textnormal{ GeV}}{(-S_{3,\textnormal{iso}})^{2/5}} f_s^{12/5}\left(\frac{\xi}{10^7\textnormal{ eV}}\right)^{6/5} \left(\frac{H_\textnormal{inf}}{10^{13}\textnormal{ GeV}}\right)^{12/5}
\end{aligned} \right\}
\left.\begin{aligned}
\textnormal{for } &f_a/N \apprle \\
&2\times10^{15}\textnormal{ GeV}.
\end{aligned}\right.
\label{htconsa}
\end{equation}
Still in the absence of entropy dilution, but for the scale independent mass (\ref{siam}) and for the axion condensing at any temperature,
\begin{equation}
\left.\begin{aligned}
	\alpha &\apprge 1.1\times10^{3}\,f_s^4\left(\frac{\zeta}{10^{-9}\textnormal{ eV}}\right) \left(\frac{H_\textnormal{inf}}{10^{13}\textnormal{ GeV}}\right)^4, \\
	-S_{3,\textnormal{iso}} &\apprge 1.2\times10^{5}\,f_s^6 \left(\frac{\zeta}{10^{-9}\textnormal{ eV}}\right)^{3/2} \left(\frac{H_\textnormal{inf}}{10^{13}\textnormal{ GeV}}\right)^6.
\end{aligned} \right.
\label{ltconsb}
\end{equation}
In the presence of entropy dilution, axions obeying respectively the scale dependent (\ref{sdam}) or scale independent (\ref{siam}) axion mass formulas have identical constraints, but with different domains of validity,
\begin{align}
\alpha \apprge .10\,f_s^4\left(\frac{H_\textnormal{inf}}{10^{13}\textnormal{ GeV}}\right)^4&, \qquad
	-S_{3,\textnormal{iso}} \apprge .093 \,f_s^6\left(\frac{H_\textnormal{inf}}{10^{13}\textnormal{ GeV}}\right)^6, \notag\\
\textnormal{for }&f_a/N \apprle 3.0\times10^{20}\textnormal{ GeV}\left(\frac{\xi}{10^{7}\textnormal{ eV}}\right)\label{dilconsa}\\
	\textnormal{or }&\zeta \apprge 3.3\times10^{-14}\textnormal{ eV}. \label{dilconsb}	
\end{align}
As mentioned in Appendix \ref{apro}, in the case of entropy dilution, but outside the domains (\ref{dilconsa}) or (\ref{dilconsb}), the above formulas are no longer valid, and the axion is constrained by (\ref{ltconsa}) or (\ref{ltconsb}).\\

\centerline{\textbf{BGW-QCD String Axion}}

The constraints listed above are for generic axions satisfying the scale dependent (\ref{sdam}) and scale independent (\ref{siam}) axion mass formulas.  Each constraint contains factors of $f_s$, where $f_s\neq 1$ means that the axion scales during inflation and condensation differ.  For the three axions we are considering, only the BGW-QCD axion has $f_s\neq 1$, as seen from (\ref{msval}).  We now construct the specific BGW-QCD constraints by plugging in its values of $\xi$ and $f_s$.

In the absence of entropy dilution and for the BGW-QCD axion condensing during $T\apprle\Lambda_\textnormal{QCD}$,
\begin{equation}
\left.\begin{aligned}
	f_a/N &\apprle (3.0\times10^{17}\textnormal{ GeV})\,\alpha^{1/3} \left(\frac{10^{13}\textnormal{ GeV}}{H_\textnormal{inf}}\right)^{4/3} \\
	f_a/N &\apprle (2.4\times10^{17}\textnormal{ GeV})\,(-S_{3,\textnormal{iso}})^{2/9} \left(\frac{10^{13}\textnormal{ GeV}}{H_\textnormal{inf}}\right)^{4/3}
\end{aligned} \right\}
\quad \textnormal{for } f_a/N \apprge 3.2\times10^{17}\textnormal{ GeV}
\label{bgwltconsa}
\end{equation}
while for the BGW-QCD axion condensing during $T\apprge\Lambda_\textnormal{QCD}$,
\begin{equation}
\left.\begin{aligned}
	f_a/N &\apprle (3.0\times10^{17}\textnormal{ GeV})\,\alpha^{3/7} \left(\frac{10^{13}\textnormal{ GeV}}{H_\textnormal{inf}}\right)^{12/7} \\
	f_a/N &\apprle (2.2\times10^{17}\textnormal{ GeV})\,(-S_{3,\textnormal{iso}})^{2/7} \left(\frac{10^{13}\textnormal{ GeV}}{H_\textnormal{inf}}\right)^{12/7}
\end{aligned} \right\}
\quad \textnormal{for } f_a/N \apprle 2\times10^{15}\textnormal{ GeV},
\label{bgwhtconsa}
\end{equation}
while in the presence of entropy dilution,
\begin{equation}
\left.\begin{aligned}
	f_a/N &\apprle (1.8\times10^{18}\textnormal{ GeV})\,\alpha^{1/4} \left(\frac{10^{13}\textnormal{ GeV}}{H_\textnormal{inf}}\right) \\
	f_a/N &\apprle (1.5\times10^{18}\textnormal{ GeV})\,(-S_{3,\textnormal{iso}})^{1/6} \left(\frac{10^{13}\textnormal{ GeV}}{H_\textnormal{inf}}\right)
\end{aligned} \right\}
\quad \textnormal{for } f_a/N \apprle 3.6\times10^{20}\textnormal{ GeV}. \label{bgwdilconsb}	
\end{equation}
As mentioned above and in Appendix \ref{apro}, in the case of entropy dilution, but outside its domain, (\ref{bgwdilconsb}) is no longer valid, and the axion is constrained by (\ref{bgwltconsa}).

\section{Conclusions}
\label{Conclusion}

The axion constraints of the previous section depend on the Hubble constant during inflation, $H_\textnormal{inf}$.  The PLANCK polarimetry experiment is expected to be sensitive to values of $H_\textnormal{inf}$ much smaller than the current bound of $H_\textnormal{inf}\apprle 2.8\times10^{14}$ GeV \cite{WMAPH}.  After briefly describing the PLANCK polarimetry experiment and its expected level of sensitivity, the constraints of the previous section will be examined under of the possibility of a precise determination of $H_\textnormal{inf}$ by PLANCK.
  
\begin{figure}[p]
	\begin{center}
		\includegraphics[width=5in]{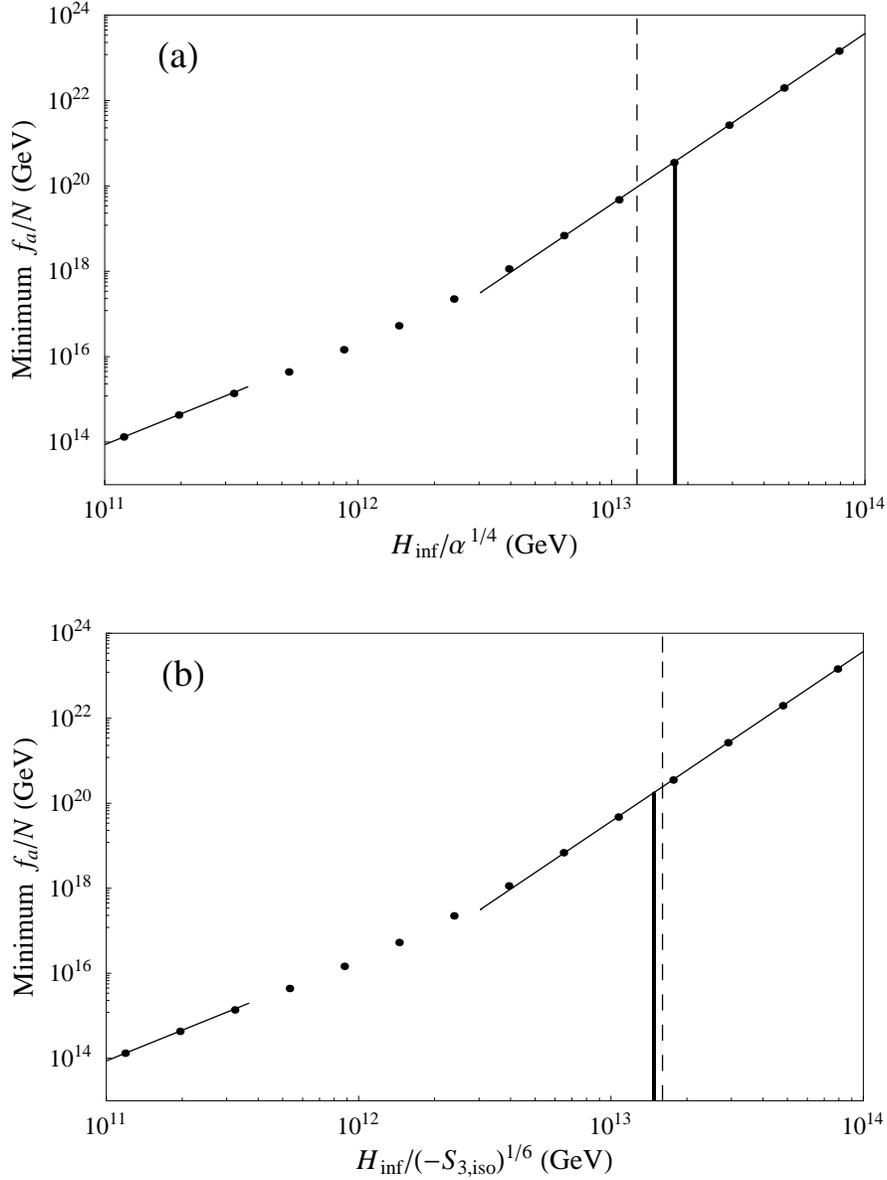}
	\end{center}
	\caption{Constraints for axions obeying the scale dependent QCD mass formula ((\ref{sdam}) with the values in (\ref{mnum})) and with the same axion scale during inflation and condensation.  This includes the model independent string axion and these plots are equivalent to the plots in \cite{fpt}.  (a) is for the isocurvature constraints while (b) is for the non-Gaussian constraints.  The diagonal lines correspond to constraints in the absence of entropy dilution, (\ref{ltconsa}) and (\ref{htconsa}).  The dots are a smooth interpolation connecting them.  The allowed parameter space is above.  The thick vertical line corresponds to constraints in the presence of entropy dilution (\ref{dilconsa}), the allowed region being to the left. The space to the right of the dashed vertical line is a reasonable estimate of the parameter space that the PLANCK polarimetry experiment will probe.}
	\label{figsdam}
\end{figure}

\begin{figure}[p]
	\begin{center}
		\includegraphics[width=5in]{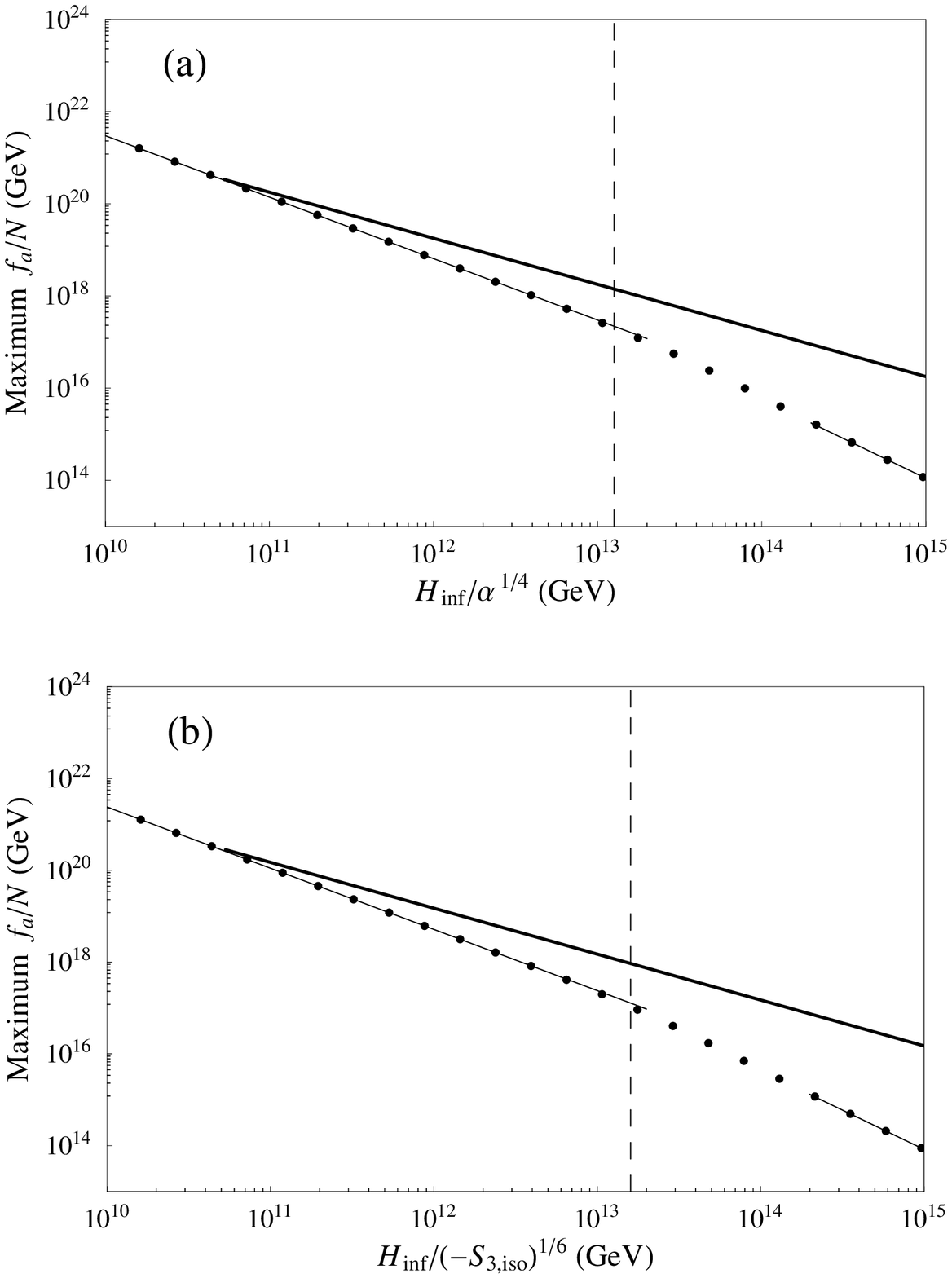}
	\end{center}
	\caption{Constraints specifically for the BGW-QCD axion.  (a) is for the isocurvature constraints while (b) is for the non-Gaussian constraints.  The thin diagonal lines correspond to constraints in the absence of entropy dilution, (\ref{bgwltconsa}) and (\ref{bgwhtconsa}).  The dots are a smooth interpolation connecting them.  The allowed parameter space is below.  The thick diagonal line corresponds to constraints in the presence of entropy dilution (\ref{bgwdilconsb}), the allowed region being below. The space to the right of the dashed vertical line is a reasonable estimate of the parameter space that the PLANCK polarimetry experiment will probe.}
	\label{figbgw}
\end{figure}

\begin{figure}[p]
	\begin{center}
		\includegraphics[width=5in]{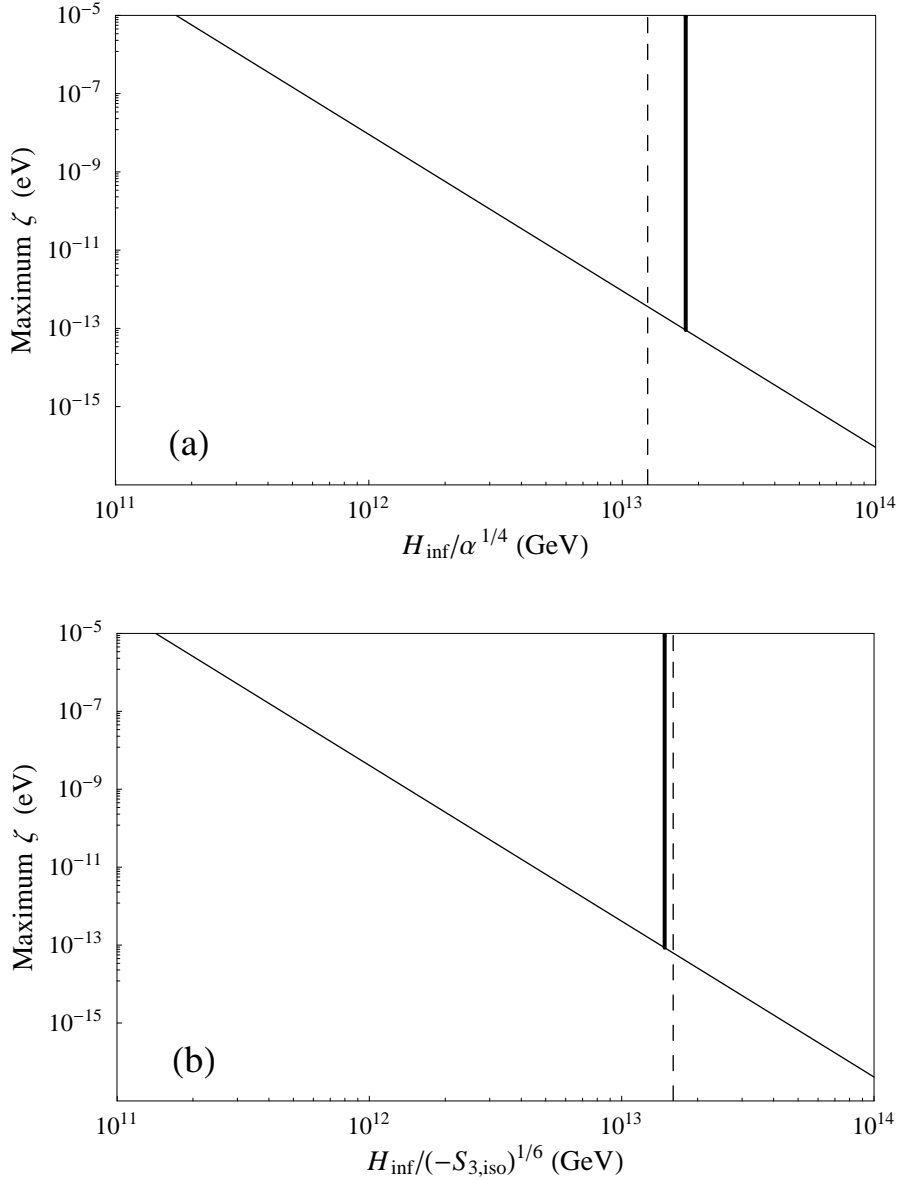}
	\end{center}
	\caption{Constraints for axions obeying the scale independent mass formula (\ref{siam}).  This includes the BGW-two condensate axion.  (a) is for the isocurvature constraints while (b) is for the non-Gaussian constraints.  The diagonal line corresponds to constraints in the absence of entropy dilution (\ref{ltconsb}), the allowed region being below.  The solid vertical line corresponds to constraints in the presence of entropy dilution (\ref{dilconsb}), the allowed region being to the left.  The space to the right of the dashed vertical line is a reasonable estimate of the parameter space that the PLANCK polarimetry experiment will probe.}
	\label{figsiam}
\end{figure}

As mentioned in section \ref{Relic} all massless fields undergo quantum fluctuations during inflation, which includes the metric.  Metric fluctuations generated during inflation appear as gravitational waves which induce tensor B-modes in CMB temperature fluctuations which PLANCK will attempt to measure.  Just as axion fluctuations can be written in terms of $H_\textnormal{inf}$, so too can metric fluctuations.  Thus, a positive measurement by PLANCK of gravitational waves interpreted as arising from inflation would be a measurement of $H_\textnormal{inf}$.  A reasonable estimate is that PLANCK will probe $H_\textnormal{inf}\apprge10^{13}$ GeV \cite{co03,fpt}.  In figures \ref{figsdam}--\ref{figsiam} the constraints of the previous section are plotted.  In these figures, the PLANCK polarimetry experiment is assumed to probe the parameter space to the right of the dashed vertical lines, which are given by
\begin{align}
	\frac{H_\textnormal{inf}}{\alpha^{1/4}}&\apprge\frac{10^{13}\textnormal{ GeV}}{(.4)^{1/4}}=1.3\times 10^{13} \textnormal{ GeV},\\ 
	\frac{H_\textnormal{inf}}{(-S_{3,\textnormal{iso}})^{1/6}}&\apprge\frac{10^{13}\textnormal{ GeV}}{(.06)^{1/6}}=1.6\times 10^{13} \textnormal{ GeV}.
\end{align}

Figure \ref{figsdam} presents the constraints (\ref{ltconsa}), (\ref{htconsa}) and (\ref{dilconsa}) for an axion obeying the scale dependent QCD mass formula (\ref{mimass}), and with $f_s=1$, which means that the axion scale during inflation and condensation is the same.  This includes the MI axion (but not the BGW-QCD axion which has $f_s\neq 1$) and these plots are equivalent to the plots in \cite{fpt}.  In the absence of entropy dilution, the allowed parameter space is above the diagonal lines (and we may ignore all other lines in the figure).  The diagonal lines do not connect because they are valid in disconnected regions.  The constraints between them should presumably be given by a smooth interpolation which connects them.  Such an interpolation is given by the dots.  The allowed parameter space, in the absence of entropy dilution, is then above the dots.  In the presence of entropy dilution we ignore the diagonal lines and dots, and the allowed parameter space is to the left of the thick vertical line.  The thick vertical line stops because the constraint it describes is only valid in the region given by (\ref{dilconsa}).  Outside this region, and thus above where the thick vertical line stops, it is the diagonal lines (and dots) which are the proper constraint, even in the presence of entropy dilution.

In the absence of entropy dilution, we can see in figure \ref{figsdam} that if PLANCK makes a positive measurement of gravitational waves interpreted as arising from inflation, and thus we are somewhere to the right of the dashed vertical line, then $f_a/N\apprge 10^{20}$ GeV (so as to be above the dots).  This rules out the existence of the MI axion, which requires, from (\ref{miscale}), $f_a/N\sim 10^{16}$ GeV.  In the presence of entropy dilution, figure \ref{figsdam}(a) shows a sliver of parameter space (the space between the vertical lines) in which a positive measurement by PLANCK could be consistent with the MI axion.  However, this possibility is ruled out in figure \ref{figsdam}(b).  Thus we find that for a positive measurement of gravitational waves interpreted as arising from inflation by PLANCK, the existence of the MI axion is ruled out \cite{fpt}.

Figure \ref{figbgw} presents the constraints (\ref{bgwltconsa})--(\ref{bgwdilconsb}) for the BGW-QCD axion.  The BGW-QCD axion obeys the scale dependent mass formula (\ref{qcdmass}) and has $f_s$ given by (\ref{msval}), which means that the axion scale during inflation differs from that during condensation.  In the absence of entropy dilution, the allowed parameter space is below the thin diagonal lines (and we may ignore all other lines in the figure).  The thin diagonal lines do not connect because they are valid in disconnected regions.  The constraints between them should presumably be given by a smooth interpolation which connects them.  Such an interpolation is given by the dots.  The allowed parameter space, in the absence of entropy dilution, is then below the dots.  In the presence of entropy dilution we ignore the thin diagonal lines and dots, and the allowed parameter space is below the thick diagonal line.  The thick diagonal line stops because the constraint it describes is only valid in the region given by (\ref{bgwdilconsb}).  Outside this region, and thus beyond where the thick diagonal line stops, it is the thin diagonal lines (and dots) which are the proper constaint, even in the presence of entropy dilution.

In the absence of entropy dilution, we can see in figure \ref{figbgw} that if PLANCK makes a positive measurement of gravitational waves interpreted as arising from inflation, and thus we are somewhere to the right of the dashed vertical line, then $f_a/N\apprle 10^{17}$ GeV (so as to be below the dots).  This rules out the existence of the BGW-QCD axion, which requires, from (\ref{bgwqcdscale}), $f_a/N\apprge 6 \times 10^{18}$ GeV.  In the presence of entropy dilution, if PLANCK makes a positive measurement then $f_a/N\apprle 10^{18}$ GeV (so as to be below the dots).  Again, the existence of the BGW-QCD axion is ruled out.  Thus we find that for a positive measurement of gravitational waves interpreted as arising from inflation by PLANCK, the existence of the BGW-QCD axion is ruled out.

Figure \ref{figsiam} presents the constraints (\ref{ltconsb}) and (\ref{dilconsb}) for an axion obeying the scale independent mass formula (\ref{siam}) and with $f_s=1$, which means that the axion scale during inflation and condensation is the same.  This includes the BGW-2C axion.  In the absence of entropy dilution the allowed parameter space is below the diagonal line (and we may ignore all other lines in the figure).  In the presence of entropy dilution we ignore the diagonal line and the allowed parameter space is to the left of the thick vertical line.  The thick vertical line stops because the constraint it describes is only valid in the region given by (\ref{dilconsb}).  Outside this region, and thus below where the thick vertical line stops, it is the diagonal line which is the proper constraint, even in the presence of entropy dilution.

In the absence of entropy dilution, we can see in figure \ref{figsiam} that if PLANCK makes a positive measurement of gravitational waves interpreted as arising from inflation, and thus we are somewhere to the right of the dashed vertical line, then $\zeta\apprle 10^{-13}$ eV (so as to be below the diagonal line).  This rules out the possibility of the BGW-2C axion obeying a scale independent mass formula, which requires, from (\ref{2cmass2}), $\zeta=m_a\apprge 10^{-9}$ eV.  In the presence of entropy dilution, figure \ref{figsiam}(a) shows a sliver of parameter space (the space between the vertical lines) in which a positive measurement by PLANCK could be consistent with the BGW-2C axion obeying a scale independent mass formula.  However, this possibility is ruled out in figure \ref{figsiam}(b).  Thus we find that for a positive measurement of gravitational waves interpreted as arising from inflation by PLANCK, the possibility of the BGW-2C axion obeying a scale independent mass formula is ruled out.

Just as in \cite{fpt} we have shown that if the PLANCK polarimetry experiment makes a positive measurement of gravitational waves interpreted as arising from inflation with a Hubble constant during inflation of $H_\textnormal{inf}\apprge 10^{13}$ GeV, then a QCD axion satisfying the usual mass formulas, (\ref{mimass}) and (\ref{htm}), must have $f_a/N\apprge10^{20}$ GeV.  In QCD, any axion scale and corresponding axion mass solves the strong CP problem, so this result has no direct bearing on the status of the QCD axion.  However, it is interesting to ask if there are theories which can supply such large axion scales.  It is well known that string axions have naturally large scales, so we have asked in this paper, following \cite{fpt}, whether two different string axions could be the QCD axion.  We have shown that if the PLANK polarimetry experiment makes a positive measurement of gravitational waves interpreted as arising from inflation, then neither the model independent axion \cite{fpt} nor the BGW-QCD axion\footnote{We note that for the BGW-QCD axion this result depends critically on the fact that the axion scale during inflation is given by (\ref{susyscale}) and not (\ref{bgwscale}).} can be the QCD axion, their existence being ruled out.  We have also shown that a third string axion, the BGW-two condensate axion, which does not solve the strong CP problem and is therefore not a candidate for the QCD axion, cannot obey a scale independent mass formula, such as (\ref{2cmass2}), if PLANCK makes a positive measurement of gravitational waves interpreted as arising from inflation.  Finally, in Appendix \ref{ltam} we have given a quantitative calculation of the corrections to the zero temperature QCD axion mass in the context of the chiral Lagrangian and justified the often held assumption that such corrections are negligible.

\section*{Acknowledgments}
I am grateful to Lotfi Boubekeur, Dan Butter, Jeff Filippini, W. Fischler, Patrick Fox, Soojin Kwon, Alexander Vilenkin and especially Mary K. Gaillard for helpful comments, corrections and discussions.  This work was supported in part by the Director, Office of Science, Office of High Energy and Nuclear Physics, Division of High Energy Physics of the U.S. Department of Energy under Contract DE-AC02-05CH11231, in part by the National Science Foundation under grant PHY-0098840.

\appendix
\section{Axion Production}
\label{apro}
In this appendix we review the production of axions in the early universe and the resulting relic axion densities.  We do so for the standard axion mass formulas as well as a scale independent formula.

There are three methods in which axions are produced in the early universe (see, for example, \cite{kt}): (1) thermal interactions, (2) decay of axionic strings, and (3) relaxation of an initially misaligned $\theta$ angle.  For axion scales, $f_a$, much above $10^{10}$ GeV, which we shall certainly be assuming, axions interact so weakly that a thermal population never results.  In an inflationary universe, which we shall also be assuming, strings axions exist before and during inflation, and so all axionic strings are inflated away.\footnote{I am thankful to A. Vilenkin for bringing a misunderstanding of mine on this issue to my attention.} This leaves only the third method \cite{pww} for axion production, which we now review.

The equation of motion for a homogeneous axion, $a$, which is related to the $\theta$ angle by $a=(f_a/N)\theta$, where $N$ is the axion anomaly coefficient, in an FRW cosmology is
\begin{equation}
	\ddot{a}+3H\dot{a}+m_a^2(T)a=0,
	\label{eom}
\end{equation}
where $H$ is the Hubble parameter and we have assumed that $a$ is small in writing the last term (which will be corrected for below).  It is important to note that the axion mass, $m_a(T)$, is temperature dependent.

At high temperatures the axion mass is very small (the axion mass formulas were presented in section \ref{Axions} and will be written again below) and the mass term in ({\ref{eom}) can be ignored.  If we assume initially that $\dot{a}=0$ then the equations of motion are solved by a constant axion field, $a_i=(f_a/N)\theta_i$, $\theta_i$ being known as the initial misalignment angle.  As the temperature falls the axion mass increases.  Once we reach the oscillation temperature (also known as the condensation temperature), $T_\textnormal{osc}$, defined by $3H(T_\textnormal{osc})=m_a(T_\textnormal{osc})$, the axion begins to oscillate around its vacuum value, and is said to have condensed.

From a look at (\ref{eom}), the energy density is given by
\begin{equation}
	\rho_a=\frac{1}{2}\dot{a}^2+\frac{1}{2}m_a^2(T)a^2.
	\label{ed}
\end{equation}
It is not hard to show that even while the axion mass is decreasing, the number density, $n_a$, scales like non-relativistic matter (see, for example, \cite{kt}).  Assuming there is no entropy dilution, this allows us to calculate the current axion number density,
\begin{equation}
	n_{a0}=\frac{\rho_{a0}}{m_a}=\frac{n_a}{s}s_0,
\end{equation}
where $m_a=m_a(T=0)$ is the zero temperature axion mass and we evaluate $n_a/s$ at the moment the axion condensate forms, so
\begin{equation}
	n_a=f_c\frac{1}{2}m_a(T_\textnormal{osc})(f_a/N)^2(T_\textnormal{osc})\left<\theta_i^2\right>f(\theta_i^2),
	\label{nd}
\end{equation}
which follows from (\ref{ed}), along with $f_c\simeq1.44$, which is a numerical correction for the temperature dependence of the axion mass and $f(\theta_i^2)$ which is a correction for anharmonic effects of the axion potential (and corrects for our assumption of small $a$ in writing the last term in (\ref{eom})),  which for small $\theta$ is roughly equal to one \cite{pww}.  $\left<\theta_i^2\right>$ is the mean square value of the initial misalignment angle, discussed in section \ref{Relic}.

The final piece of information that we need is the axion mass.  Axion masses for the three string axions we are considering were presented in section \ref{Axions}.  For two of the axions (the MI and BGW-QCD axions) the zero temperature masses were inversely proportional to the axion scale, while for the third axion (the BGW-2C axion) the mass was independent of the axion scale.  Consequently, we will derive relic axion densities for the following two mass formulas,
\begin{align}
	m_a&=\xi \left(\frac{1\textnormal{ GeV}}{f_a/N}\right),\label{sdamap}\\
	m_a&=\zeta,\label{siamap}
\end{align}
which we will refer to as the scale dependent and scale independent mass formulas, and where $\xi$ and $\zeta$ are constants (which have dimensions of mass).  The scale dependent mass (\ref{sdamap}) will be used for temperatures $T\apprle \Lambda_{\textnormal{QCD}}$, which we justify in Appendix \ref{ltam}, while the scale independent mass (\ref{siamap}) will be used for all temperatures, which we justify below.  For axions which obey the scale dependent formula (\ref{sdamap}), but condense at temperatures $T\apprge\Lambda_{\textnormal{QCD}}$, we will need the high temperature axion mass formula, which may be calculated by considering instantons at finite temperature \cite{gpy}.  One finds \cite{pww}
\begin{equation}
	m_a(T) \simeq 2.2\times10^5 \textnormal{ eV} \left(\frac{1\textnormal{ GeV}}{f_a/N}\right) \left(\frac{\Lambda_\textnormal{QCD}}{200\textnormal{ MeV}}\right)^{1/2}\left(\frac{\Lambda_\textnormal{QCD}}{T}\right)^4,
	\label{tm}
\end{equation}
which is accurate for $T\apprge\Lambda_{\textnormal{QCD}}$, but not for too much larger.

With these mass formulas, $3H=m_a(T_\textnormal{osc})$, the Friedman equation, $\rho=3H^2M_p^2$, and the thermal energy density in a radiation dominated era, $\rho=(\pi^2/30)g_* T^4_\textnormal{osc}$, where $g_*=61.75$ for $T_\textnormal{osc}$ just above $\Lambda_\textnormal{QCD}$ and $g_*=10.75$ for $T_\textnormal{osc}$ just below $\Lambda_\textnormal{QCD}$, we may solve for the oscillation temperature.  For axions obeying the scale dependent mass formula (\ref{sdamap}) which condense, respectively, during $T_\textnormal{osc} \apprle \Lambda_\textnormal{QCD}$ or $T_\textnormal{osc} \apprge \Lambda_\textnormal{QCD}$,
\begin{align}
	T_\textnormal{osc} &\simeq 3.3\times10^{7}\textnormal{ MeV}\left(\frac{\xi}{1\textnormal{ eV}}\right)^{1/2} \left(\frac{1\textnormal{ GeV}}{f_a/N}\right)^{1/2}, \label{ltosca}\\
	T_\textnormal{osc} &\simeq 7.0\times10^{4}\textnormal{ MeV } \left(\frac{\Lambda_\textnormal{QCD}}{200\textnormal{ MeV}}\right)^{3/4} \left(\frac{1\textnormal{ GeV}}{f_a/N}\right)^{1/6}, \label{htosc}
\end{align}
while for axions which obey the scale independent mass formula (\ref{siamap}) and condense at any temperature,\footnote{$g_*$ depends on the value of $T_\textnormal{osc}$ and in this sense this formula is not valid for all temperatures.  However, $T_\textnormal{osc}\sim g_*^{-1/4}$ and the error in using $g_*= 75.75$ (which we use in this and future equations for the scale independent mass (\ref{siamap})) for all oscillation temperatures is small.}
\begin{equation}
	T_\textnormal{osc} \simeq 2\times10^{7}\textnormal{ MeV}\left(\frac{\zeta}{1\textnormal{ eV}}\right)^{1/2}. \label{ltoscb}
\end{equation}

We may now form the relic axion density, $\Omega_a=\rho_{a0}/\rho_c$, where $\rho_c=2h_0^2M_p^2$ is the critical density and $M_p=2.4\times10^{18}$ GeV is the reduced Planck mass.  Using $s=(2\pi^2/45)g_{*s} T^4_{\textnormal{osc}}$, $g_{*s}=61.75$ for $T_\textnormal{osc}$ just above $\Lambda_\textnormal{QCD}$ and $g_{*s}=10.75$ for $T_\textnormal{osc}$ just below $\Lambda_\textnormal{QCD}$, with $T_{\textnormal{osc}}$ given above, and $s_0=(2\pi^2/45)g_{*s_0} T_0^4$, $g_{*s0}=3.91$, $T_0=2.73$ K, we have for axions obeying the scale dependent mass formula (\ref{sdamap}) which condense, respectively, during $T_\textnormal{osc} \apprle \Lambda_\textnormal{QCD}$ or $T_\textnormal{osc} \apprge \Lambda_\textnormal{QCD}$,
\begin{align}
	\Omega_a h^2 &\simeq 1.8\times 10^{-24} \left(\frac{\xi}{1\textnormal{ eV}}\right)^{1/2} \left(\frac{f_a/N}{1 \textnormal{ GeV}}\right)^{3/2} \left<\theta_i^2\right> f(\theta_i)^2, \label{aprelicla}\\
	\Omega_a h^2 &\simeq 3.6\times 10^{-22} \left(\frac{\xi}{1\textnormal{ eV}}\right)\left(\frac{f_a/N}{1 \textnormal{ GeV}}\right)^{7/6} \left<\theta_i^2\right> f(\theta_i)^2, \label{aprelicha}
\end{align}
while for axions which obey the scale independent mass formula (\ref{siamap}) and condense at any temperature,
\begin{equation}
\Omega_a h^2 \simeq 1.1\times 10^{-24} \left(\frac{\zeta}{1\textnormal{ eV}}\right)^{1/2} \left(\frac{f_a/N}{1 \textnormal{ GeV}}\right)^2 \left<\theta_i^2\right> f(\theta_i)^2, \label{apreliclb}
\end{equation}
where $h=H_0/(100\textnormal{ km s}^{-1}\textnormal{ Mpc}^{-1})$ and we have taken $\Lambda_\textnormal{QCD}=200$ MeV.  Using (\ref{ltosca}) and (\ref{htosc}) the domains of validity for (\ref{aprelicla}) and (\ref{aprelicha}) may be rewritten as
\begin{align}
	T_\textnormal{osc} \apprle \Lambda_\textnormal{QCD} \quad&\Leftrightarrow\quad f_a/N \apprge 2.7\times10^{10}\textnormal{ GeV} \left(\frac{\xi}{1\textnormal{ eV}}\right),\label{ltfa}\\
	T_\textnormal{osc}\apprge\Lambda_\textnormal{QCD} \quad&\Leftrightarrow\quad f_a/N \apprle 2\times10^{15}\textnormal{ GeV} \label{htf},
\end{align}
where again we have taken $\Lambda_\textnormal{QCD}=200$ MeV.  

In section \ref{sum} the numerical values for the axion masses and scales we will be using in this paper were given.  With these values, (\ref{ltfa}) and (\ref{htf}) tell us that for axions obeying the scale dependent mass formula (\ref{sdamap}), $T_\textnormal{osc}\sim\Lambda_\textnormal{QCD}$ and below, and that in some cases we are in fact considering oscillation temperatures where neither the high temperature nor the zero temperature mass formulas are accurate.  Presumably, in these cases, the correct relic axion density is given by the smooth interpolation between the densities derived from high and low temperature mass formulas.  We also see that we will never be considering oscillation temperatures above a TeV.  One of the axions we are considering (the BGW-QCD axion) has its mass and scale modified above the scale of supersymmetry breaking (see section \ref{bgwqcd}).  If we take this scale to be $\sim$ TeV or above then we will not have to take this into account here (we will have to take this in account in section \ref{Relic} when we consider axion fluctuations generated during inflation).  For the axion obeying the scale independent mass formula (\ref{siamap}), we find that $T_\textnormal{osc}\sim\Lambda_2$, where above $\Lambda_2$ the scale independent mass formula is no longer valid.  As mentioned in section \ref{bgw2c} we will assume the scale independent mass formula is valid for all oscillation temperatures since otherwise the cosmological consequences of this axion would not differ much from the others.

So far everything has been derived in the absence of entropy dilution.  We now consider the possibly of a late entropy release \cite{lsss}, diluting the axion condensate by the late decay of a massive, non-relativistic particle.  For the late decaying particle to dilute the axion condensate it must decay after the axion condensate is formed, and therefore have a reheating temperature satisfying $T_\textnormal{RH}\apprle T_\textnormal{osc}$.  However, any entropy release is constrained by the success of big bang nucleosynthesis, which in this case requires $T_\textnormal{RH}\apprge6$ MeV \cite{rs}.  To calculate the relic axion density we must form $n_a/s$ after the particle has decayed.  The result is \cite{kmy}
\begin{equation}
	\Omega_a h^2 \simeq 3.3\times10^{-31}\left(\frac{T_\textnormal{RH}}{6\textnormal{ MeV}}\right) \left(\frac{f_a/N}{1\textnormal{ GeV}}\right)^2 \left<\theta_i^2\right>f(\theta_i)^2.
	\label{apdil}
\end{equation}
Using (\ref{ltosca}) and (\ref{ltoscb}) we may rewrite the domain of (\ref{apdil}), $T_\textnormal{osc}\apprge T_\textnormal{RH}$, for axions obeying, respectively, the scale dependent (\ref{sdamap}) and scale independent (\ref{siamap}) mass formulas, as
\begin{align}
	f_a/N &\apprle 3.0\times10^{13}\textnormal{ GeV}\left(\frac{\xi}{1\textnormal{ eV}}\right)\left(\frac{6\textnormal{ MeV}}{T_\textnormal{RH}}\right)^2, \label{dilfa}\\
	\zeta &\apprge 3.3\times10^{-14}\textnormal{ eV}\left(\frac{T_\textnormal{RH}}{6\textnormal{ MeV}}\right)^2.\label{dilfb}
\end{align}
Outside these domains, the axion condenses after the particle decays, which is during a radiation dominated era, and therefore the axion density is given by (\ref{aprelicla}) or (\ref{apreliclb}).

\section{Low Temperature QCD Axion Mass}
\label{ltam}
In this appendix we explicitly calculate temperature corrections to the zero temperature QCD axion mass using the chiral Lagrangian.  This will quantitatively justify the use of the zero temperature mass for temperatures $T\apprle\Lambda_\textnormal{QCD}$.

For two light quarks, the chiral Lagrangian, derived from considerations of chiral symmetry breaking, is a low energy effective theory of pions, and to lowest order (apart from counterterms) is given by \cite{we79}
\begin{equation}
	{\cal L}=\frac{f^2_\pi}{4}\textnormal{Tr}\left(\partial_\mu U^\dag \partial^\mu U\right) +\frac{f^2_\pi}{4}B_0\textnormal{Tr}\left(MU^\dag +UM^\dag\right),
	\label{chilag}
\end{equation}
where $f_\pi=96$ MeV is the pion decay constant, $B_0$ is a constant describing the scale at which chiral symmetry breaking ocurrs, $M=\textnormal{diag}(m_u,m_d)$ is the two quark mass matrix and
\begin{equation}
	U=e^{i\boldsymbol{\pi}/f_\pi},\qquad \boldsymbol{\pi}=\pi_i\sigma_i, \quad i=1,2,3,
	\label{pipar}
\end{equation}
where $\sigma_i$ are the Pauli matrices, $\pi_i$ are the pions and there is a sum over $i$.  This is the lowest order Langrangian and will be all that we will consider here.  It is the unique Lagrangian satisfying the required symmetries and of dimension two, where a derivative has dimension one and the mass matrix has dimension two.  There are higher order Lagrangians, the next one being of dimension four \cite{gl85}.  This Lagrangian is only valid for energies where QCD is confined, and can therefore form bound states such as pions.  Thus, formulas derived from the chiral Lagrangian are taken to be valid only for energies and temperatures below $\Lambda_\textnormal{QCD}$.

To incorporate the axion \cite{weqft2}, the mass matrix is modified by
\begin{equation}
	M\rightarrow e^{ia/f_a}M,
\end{equation}
where $a$ is the axion and $f_a$ is the axion scale.  Plugging this and (\ref{pipar}) into the chiral Lagrangian (\ref{chilag}) and expanding one finds the axion mixing with $\pi_3$, which, if electric charge is incorporated, is the neutral pion.  Diagonalizing the mass matrix leads to the usual pion mass, plus $O(f_\pi/f_a)^2$ corrections, and the zero temperature QCD axion mass,
\begin{equation}
	m^2_a=\left(\frac{f_\pi}{f_a}\right)^2\frac{m_u m_d}{(m_u+m_d)^2}m_\pi^2+O(f_\pi/f_a)^4,
\end{equation}
where $m_u$ and $m_d$ are the up and down quark masses and $m_\pi$ is the neutral pion mass.

We will calculate temperature corrections using the partition function \cite{gl87}.  In particular we will calculate the one loop correction to the scalar potential at finite temperature \cite{we74}.  The partition function requires moving to imaginary time, integrating imaginary time from $0$ to $\beta=1/T$, where $T$ is the temperature, and taking all fields to be periodic in imaginary time with period $\beta$.

To evaluate the path integral we make the usual first step of shifting the fields by their classical solutions.  We could expand the chiral Lagrangian (\ref{chilag}) first and then make this shift, but it proves easier to parametrize the shift as follows \cite{dgh},
\begin{equation}
	U= \bar{U}e^{i\boldsymbol{\Delta}/F},\qquad a= \bar{a}+\delta,\qquad\boldsymbol{\Delta}=\Delta_i\sigma_i, \quad i=1,2,3,	
\end{equation}
where $\bar{U}$ and $\bar{a}$ are the classical fields (i.e. the solutions to the equations of motion) and $\Delta_i$ and $\delta$ are the dynamical fields to be integrated over in the partition function.

The partition function leads to an integral that must be done numerically, and is part of the one loop corrected mass matrix which must be diagonalized.  Upon doing so, the axion mass is obtained and can be plotted as a function of temperature, which is the top line in figure \ref{figamass}.\begin{figure}
	\begin{center}
		\includegraphics[width=5in]{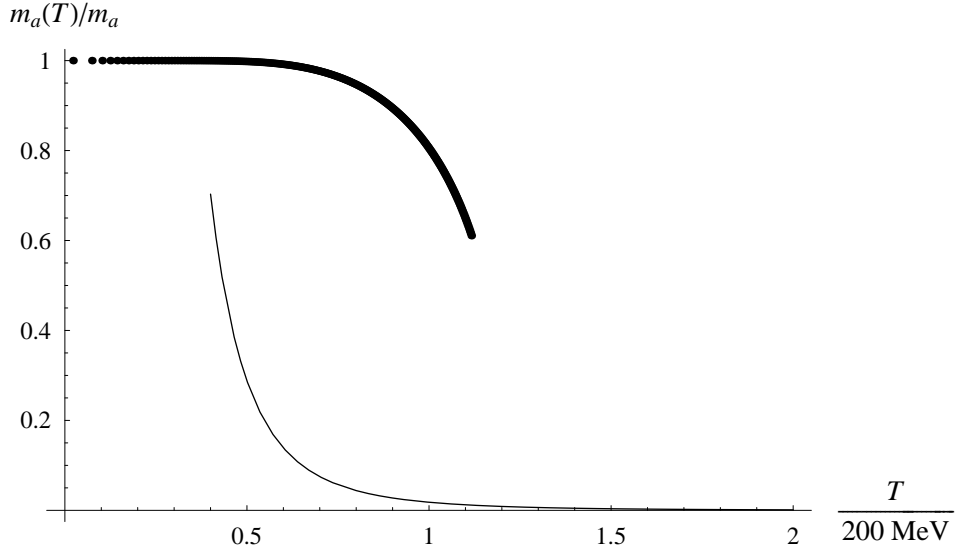}
	\end{center}
	\caption{Both curves are plots of the finite temperature QCD axion mass divided by the zero temperature mass.  The top line is only valid for $T\apprle\Lambda_\textnormal{QCD}$, while the bottom line is only valid for $T\apprge\Lambda_\textnormal{QCD}$, although each is plotted outside their respective domain.}
	\label{figamass}
\end{figure}  For the purpose of comparison, also included in the plot as the bottom line is the high temperature axion mass, (\ref{htm}) or (\ref{tm}), which is derived from (unconfined) QCD and therefore is only valid for temperatures greater than $\Lambda_\textnormal{QCD}$.

In terms of orders of magnitude, if we take $\Lambda_\textnormal{QCD}\sim O(10^2)$ MeV, then the top line will be valid up to around $T\sim O(10)$ MeV, where we can see that the axion mass barely changes.  Where the top line does begin to curve appreciatively, say at $T\simeq200$ MeV, we can no longer trust the plot, as the chiral Lagrangian, from which it is derived, is no longer valid.

Figure \ref{figamass} shows that we may take the low temperature axion mass to be equal to its zero temperature value for temperatures $T\apprle\Lambda_\textnormal{QCD}$.

\end{document}